\documentstyle[aps,epsf,rotate,multicol]{revtex}

\begin{document}

\draft

\title{Path Integral Approach to Strongly Nonlinear Composites}

\author{Marc Barth\'el\'emy}

\address{CEA-Centre d'Etudes de Bruy\`eres-le-Ch\^atel\\
Service de Physique de la Mati\`ere Condens\'ee\\
BP12, 91680 Bruy\`eres-Le-Ch\^atel\\
France
}

\maketitle

\begin{abstract}
We study strongly nonlinear disordered media using a functional
method. We solve exactly the problem of a nonlinear impurity in a
linear host and we obtain a Bruggeman-like formula for the effective
nonlinear susceptibility. This formula reduces to the usual Bruggeman
effective medium approximation in the linear case and has the
following features: (i) It reproduces the weak contrast expansion to
the second order and (ii) the effective medium exponent near the
percolation threshold are $s=1$, $t=1+\kappa$, where $\kappa$ is the
nonlinearity exponent. Finally, we give analytical expressions for
previously numerically calculated quantities.
\end{abstract}



\begin{multicols}{2}



The study of the properties of linear heterogeneous media (composites,
suspensions) has been the subject of an intense activity for already
fifty years (see the reviews \onlinecite{Landauer78} and
\onlinecite{BergmanStroud92}). More recently there has been a great
interest in non-linear media \onlinecite{BergmanStroud92}. The
nonlinear composites are important for technology and also
from a fundamental point of view, for which it is important and
challenging to understand the interplay between nonlinearity and
disorder. There are essentially two types of nonlinear media. (a) Weak
nonlinearity: the nonlinearity is small compared to the linear
term. This case was studied by many authors and is now relatively well
understood\cite{StroudHui88,Zeng88,Zeng89,Tremblay92,Ponte92,Zhang94,Levy94,HuiWan95,Pellegrini00}.(b)
Strong nonlinearity: the nonlinearity is here the dominant term and
this can happen in essentially two situations. First, there can be a
sharp threshold between two different behaviors and this case can
model the phenomena of fracture or dielectric
breakdown\cite{Chakrabarti88,Roux90}. Second, the constitutive law can
be a pure power law of the form
\begin{equation}
\label{j-enonlin}
j=\chi |E|^{\kappa}E
\end{equation}
where $j$ is the current and $E$ the electric field. This behavior can
be observed in a dielectric illuminated by a
Laser\cite{BergmanStroud92} when the multiphoton processes dominates
and the usual linear approximation completely breaks down. Certain
cermets resistors, ZnO based varistors\cite{Einziger87,Niklasson89} or
disordered alloys\cite{Osofsky88} can also display this behavior
(\ref{j-enonlin}).

In a disordered medium, the nonlinear susceptibilities $\chi$ can
fluctuate from point to point and one is interested in the macroscopic
effective behavior of such a medium. If the nonlinearity exponent
$\kappa$ is the same for all phases of the medium, then the effective
nonlinear susceptibility is well-defined and is given by
$j_0=\chi_e|E_0|^{\kappa}E_0$ where $j_0$ and $E_0$ are the
macroscopic current and electric field respectively. It is difficult
to evaluate $\chi_e$ and devising a reliable method to compute
$\chi_e$ would allow one to study a variety of other problems such as
fracture or dielectric breakdown.

In this strongly nonlinear case, Blumenfeld and Bergman obtained the
weak contrast expansion to second order\cite{Blumenfeld89}. This
expansion was recovered by means of a path integral
method\cite{Barthelemy98}. The dilute limit was studied
in\cite{Levy92,Hui96}. Problems arise when one tries to find an
effective medium approximation (EMA) for this type of media. A good
EMA should satisfy the two following criteria. (i) It should
reproduces the weak-contrast expansion (at least up to the second order)
and the dilute limit (although this last condition is probably very
difficult to fullfill for strongly nonlinear media). (ii) Close to the
percolation threshold $p_c$, the effective nonlinear susceptibility
$\chi_e$ is described by two exponents: For a metal/insulator mixture,
one has\cite{Straley84,Bergman91}
\begin{equation}
\chi_e\sim (p-p_c)^{t(\kappa)}
\end{equation}
where $p$ is the proportion of the conducting component. For a
superconductor/metal mixture, one expects
\begin{equation}
\chi_e\sim (p_c-p)^{-s(\kappa)}
\end{equation}
where $p$ is here the proportion of the superconducting
component. Differents values for these exponents were proposed.
In\cite{Bergman91}, the effective medium values are
$t(\kappa)=1+\kappa$ and $s(\kappa)=1$, and
in\cite{Ponte92b,Wan96}, the exponents are
$t(\kappa)=s(\kappa)=1+\kappa/2$. In both cases, the crossover
exponent $\phi=s+t$ is equal to $2+\kappa$. These two sets of
exponents satisfy the duality relation for $d=2$\cite{Straley84}:
$t(\gamma)=\gamma s(1/\gamma)$ where $\gamma=1+\kappa$. So far,
numerical results\cite{LeeSiu95} and series analysis\cite{Meir86}
suggests that for $d=2$ the exponents $s$ and $t$ are different,
ruling out $s=t=1+\kappa/2$ although further numerical studies are
necessary to make a definitive statement. An acceptable EMA should
predict such kind of values.

We can distinguish two differents classes of approaches to this
problem.  A first
approach\cite{Ponte92,Wan96,HuiWoo95,Gao96,Hui97,Ponte97,Ponte98}
consists in expressing the effective nonlinear susceptibility in terms
of the averaged electric field in each
component. In\cite{Wan96,HuiWoo95,Gao96,Hui97}, a kind of a
``decoupling approximation'' is proposed for calculating these fields,
and one obtains a set of coupled equations which is solved
numerically. Although the agreement with numerical simulations is
generally fairly good, there are a few drawbacks to this method. In
particular, this method relies quite heavily on numerics and it is
difficult to check some analytical properties. Moreover, the
weak-contrast expansion (condition (i)) is usually not recovered and
the exponents are difficult to estimate. In particular, the mean-field
theory proposed in\cite{Wan96} does not reproduce the weak contrast
expansion but instead the lower bound established by Ponte Casta\~neda et
al\cite{Ponte92}. In this case\cite{Wan96}, the values of the
exponents are $s=t=\kappa/2+1$. In another series of
papers\cite{Ponte97,Ponte98}, the nonlinear host is linearized
up to the second order and the local electric fields are computed in a
self-consistent way. With this method, Ponte Casta\~neda and
Kailasam\cite{Ponte97} proposed an effective medium approximation
which reproduces the weak contrast expansion, but for which the
exponents are difficult to estimate.

The second approach is in the Bruggeman spirit\cite{Bruggeman35} and
consists essentially in considering an impurity in an effective
host. Bruggeman's theory was reformulated in order to apply to this
problem\cite{Bergman91} and it was further investigated by different
authors\cite{LeeSiu95,LeeYuen95,LeeYu95,Sali97}. This approach
predicts the effective medium values $s=1$ and $t=1+\kappa$ and
reproduces the weak contrast expansion to second order. However, all
the studies so far are mostly numerical and we propose here the
analytical solution. The obtained formula is relatively simple and
satisfies conditions (i) and (ii). We give the analytical expressions
for numerically estimated
quantities\cite{LeeSiu95,LeeYuen95,Sali97}. The only drawback of our
result is that the percolation threshold depends on $\kappa$ (in the
same way as in\cite{Sali97}). In the discussion, we address this point
and propose a possible way to correct this wrong behavior.


The constitutive relation is $j=\chi(r)|E|^{\kappa}E$ and the local
energy density associated to it is
\begin{equation}
w(r,E(r))=\frac{\chi (r)}{\kappa+2}|E|^{\kappa+2}
\end{equation}
$E$ is the applied field and for a heterogeneous medium the quantitie
$\chi(r)$ at point $r$ is distributed according to a binary law (a
generalization of our method to other types of disorder should be
without problems)
\begin{equation}
\label{pchi}
P(\chi=\chi(r))=p\delta (\chi-\chi_1)+q\delta (\chi-\chi_2)
\end{equation}
The total dissipated energy is given by
$W^*=\chi_e|E_0|^{\kappa+2}/(\kappa+2)$, and can be expressed as a
constrained minimum
\begin{equation}
\label{mini}
W^*=\langle\text{min}_{\bar{E}=E_0,E=-\nabla\phi}\int d^dr w(r,E(r))\rangle
\end{equation}
Here, we have assumed that $W^*$ is a self-averaging quantity in the
thermodynamic limit, which allows us to compute the average over the
disorder (the brackets $\langle\cdot\rangle$ denote the average over
disorder or equivalently, the spatial average). The minimum in
Eq.~(\ref{mini}) can be written with the help of path
integrals\cite{Barthelemy98}
\begin{equation}
\label{beta-infini}
 W^*=\text{lim}_{\beta\rightarrow\infty}
-\frac{1}{\beta}\int\tilde{\cal D}Ee^{-\beta{\cal H}}
\end{equation}
where the ``Hamiltonian'' is ${\cal H}=\int d^dr w(r,E(r))$ and where
the measure is ${\tilde{\cal D}}E={\cal D}(E,\phi)\delta ({\bar
E}-E_0)\delta(E+\nabla\phi)$. The important quantity to study is thus
the `partition function'
\begin{equation}
\label{parti-func}
Z=\int\tilde{\cal D}Ee^{-\beta{\cal H}}
\end{equation}

In a preceeding paper\cite{Barthelemy98}, we made a perturbation
expansion up to the second order in disorder and we recovered known
results\cite{Blumenfeld89}. We also showed in another
paper\cite{PellegriniBarthelemy00} how to recover Bruggeman's
approximation in the functional framework and we recall briefly the
idea. We start from the expression (\ref{parti-func}) and we add and
substract a Gaussian ansatz ${\cal H}_0=\int d^dr w_0(E(r))$
\begin{equation} 
Z=\int\tilde{\cal D}Ee^{-\beta{\cal H}_0}e^{-\beta({\cal H}-{\cal H}_0)}
\end{equation}
We expand the second exponential and resum it keeping only the
contribution at the same point
\begin{eqnarray}
\nonumber
e^{-\beta({\cal H}-{\cal H}_0)}&=&\sum_{k=1}^{\infty}\left[\int d^dr
(w(r,E(r))-w_0(E(r)))\right]^k\\
\nonumber
&\simeq & \int
d^dr\sum_{k=1}^{\infty}\left[w(r,E(r))-w_0(E(r))\right]^k\\
&\simeq & 
\int d^dr e^{-\beta [w(r,E(r))-w_0(E(r))]}
\end{eqnarray}
(here and in the following, we omit unimportant volume factors
and cut-offs).  The partition function $Z$ is thus given by
\begin{equation} 
\label{zfunc}
Z\simeq\int d^dr\int\tilde{\cal D}Ee^{-\beta{\cal H}_0}
e^{-\beta\left[w(r,E(r))-w_0(E(r))\right]}
\end{equation}
The following physical picture can be associated with this
approximation. The background is described by ${\cal H}_0$ and at
point $r$ there is an impurity described by $w-w_0$. The ideal case
would be to take a nonlinear background described by an effective
nonlinear susceptibility ${\cal
H}_0=\frac{\chi_e}{\kappa+2}\int|E|^{\kappa+2}$ and a nonlinear
impurity, which is so far impossible to compute. We thus have to
resort to a further approximation. The averaged value of the electric
field is fixed and given by $\bar{E}=E_0$. It is thus reasonable to
assume that the electric field in the background will not fluctuate
too much (at least far from the impurity) and we can expand the
nonlinear background around $E(r)=E_0$. We write
$E(r)=E_0+\varepsilon(r)$ and expand up to the second order in
$\varepsilon$
\begin{eqnarray}
\label{w0approx}
\nonumber w_0\simeq
\frac{\chi_e}{\kappa+2}E_0^{\kappa+2}+E_0^{\kappa}E_0\cdot\varepsilon(r)\\
+\frac{1}{2}E_0^{\kappa}
\sum_{i,j}\varepsilon_i(r)(\delta_{ij}+\kappa\frac{E_{0i}E_{0j}}{E_0^2})
\varepsilon_j(r)
\end{eqnarray}
while we keep the exact expression for
$w=\frac{\chi}{\kappa+2}|E|^{\kappa+2}$.  The final picture is then
the following: we compute exactly the perturbation induced by a
nonlinear impurity in the nonlinear {\it linearized} effective
medium. In addition to allow calculations, this scheme ensures that
the weak-contrast expansion will be recovered. The numerical study of
this problem can be found in \cite{LeeSiu95,LeeYuen95,Sali97}.

The path integral (\ref{zfunc}) together with the approximation
(\ref{w0approx}) can be computed and after some calculations, one is
led to
\begin{equation}
\label{zcol}
Z\simeq e^{-\beta\chi_eE_0^{\kappa+2}}\int d^due^{-\beta {\cal M}(u)}
\end{equation}
with
\begin{eqnarray}
\nonumber
{\cal M}(u)=\frac{\chi(r)}{\kappa+2}u^{\kappa+2}+\chi_eE_0^{\kappa}u\cdot
E_0\\
+\frac{\chi_eE_0^{\kappa}}{2}(E_0+u)\cdot(1-\frac{1}{I}+\kappa\frac{E_0\otimes
E_0}{E_0^2})\cdot(E_0+u)
\end{eqnarray}
where $u$ is a $d$-dimensional vector. The quantity $I$ is given
by
\begin{equation}
I(d,\kappa)=\frac{S_{d-1}}{S_d}\int_0^\pi
d\theta\sin^{d-2}\theta\frac{\cos^2\theta}{1+\kappa\cos^2\theta}
\end{equation}
where $S_d=2\pi^d/\Gamma(d/2)$ is the surface of the $d$-dimensional
sphere. In the linear case, $I(d,\kappa=0)=1/d$, and for
$d=2,3$, one has\cite{Blumenfeld89,Barthelemy98}
\begin{eqnarray}
I(2,\kappa)&=&\frac{1}{\kappa}\left(1-\frac{1}{\sqrt{1+\kappa}}\right)\\
I(3,\kappa)&=&\frac{1}{\kappa}\left(1-\frac{1}{\sqrt{\kappa}}\arcsin
\sqrt{\frac{\kappa}{1+\kappa}}\right)
\end{eqnarray}
We are interested in the large $\beta$ limit [Eq.~(\ref{beta-infini})]
so we can apply the saddle-point method to the integral
(\ref{zcol}). The saddle point $u_*$ is parallel to $E_0$ and is given
by
\begin{equation}
\label{v-def}
u_*(x)=\frac{-E_0(\kappa-\frac{1}{I})}{1-\frac{1}{I}+\kappa-
\frac{u_*^\kappa}{x E_0^\kappa}}
\end{equation}
where here $x=\chi_e/\chi$. One can note that $u_*$ is the electric
field in the nonlinear impurity embedded in the linearized effective
homogeneous host. The effective energy $W^*=-\frac{1}{\beta}\langle\ln
Z\rangle$ is then given by $W^*=W_0+\Delta W$ where
$W_0=\frac{\chi_e}{\kappa+2}E_0^{\kappa+2}$.  The natural
self-consistent condition $\Delta W=0$ can be rewritten for the
binary disorder (\ref{pchi}) under the form\cite{Bergman91,Sali97}
\begin{equation}
\label{funcbrugdef}
pf(\frac{\chi_e}{\chi_1})+qf(\frac{\chi_e}{\chi_2})=0
\end{equation}
where
\begin{eqnarray}
\label{funcbrug}
\nonumber
f(x)=\frac{1}{x}\frac{v_*^{\kappa+2}}{\kappa+2}-\frac{1}{\kappa+2}+1+v_*(x)\\
-\frac{1}{2}[1+v_*(x)]^2(1-\frac{1}{I}+\kappa)
\end{eqnarray}
where $v_*(x)=u_*(x)/E_0$ is given by Eq. (\ref{v-def}). This equation
(\ref{funcbrugdef}) [together with (\ref{v-def}) and (\ref{funcbrug})]
is our main result, and we will now discuss it.


In the linear case ($\kappa=0$), one can easily check that (\ref{funcbrugdef})
reduces to Bruggeman's\cite{Bruggeman35} equation. Moreover, in the
one-dimensional case, one recovers the exact result $1/\chi_e=\langle
1/\chi^{\frac{1}{\kappa+1}}\rangle^{\kappa+1}$.

As expected, the weak contrast expansion is recovered up to the second
order, namely
\begin{equation}
\chi_e\simeq
\langle\chi\rangle-\frac{\kappa+2}{2\langle\chi\rangle}
\langle\delta\chi^2\rangle I(d,\kappa)
\end{equation}
This fact is not surprising since we used as an ansatz ${\cal H}_0$ the
nonlinear effective medium linearized up to second order.

Our approximation will not reproduce the exact dilute limit, since the
nonlinear host is linearized. Instead, we will obtain the following
expansion to the first order in concentration (exact for $\kappa=0$)
\begin{equation}
\chi_e\simeq\chi_1+q(\kappa+2)\chi_1f(\frac{\chi_1}{\chi_2})
\end{equation}
where $q$ is the fraction of component $\chi_2$.

The critical behavior is determined by $f(x)$ for $x\simeq 0$ and
$x\to\infty$\cite{Bergman91,Sali97}. One obtains from
Eqs. (\ref{v-def},\ref{funcbrug}) $f(x\simeq 0)\simeq
f(0)-a(\kappa+1)x^{\frac{1}{\kappa+1}}$ where
\begin{eqnarray}
\nonumber 
f(0)&=&-\frac{1}{\kappa+2}+\frac{1}{2}+\frac{1}{2I}-\frac{\kappa}{2}\\
a&=&\frac{1}{\kappa+2}(-\frac{1}{I}+\kappa)^{\frac{\kappa+2}{\kappa+1}}
\end{eqnarray}
For $x\to\infty$, one has $f(x)\simeq f(\infty)+\frac{b}{x}$ where
\begin{eqnarray}
\nonumber\\
f(\infty)&=&-\frac{1}{\kappa+2}+1-\frac{1+2(-\frac{1}{I}+\kappa)}{2(1-\frac{1}{I}+\kappa)}\\
b&=&\frac{1}{\kappa+2}(\frac{\frac{1}{I}-\kappa}{1-\frac{1}{I}+\kappa})^{\kappa+2}
\end{eqnarray}
For $\kappa=0$, one recovers the known exact
expressions\cite{Bergman91,Sali97}: $a=d^2$, $b=[d/(d-1)]^2$, $f(0)=d$
and $f(\infty)=-d/[2(d-1)]$ (the function $f$ is defined up to a
constant factor, and there is a global additional factor
$1/(\kappa+2)$ in our result). These four different
coefficients were estimated numerically\cite{LeeSiu95,LeeYuen95,Sali97}.

The percolation threshold is given by 
\begin{eqnarray}
p_c=\frac{f(\infty)}{f(\infty)-f(0)}=\frac{1}{\kappa+2}\frac{\frac{2}{I}-\kappa}{(-\frac{1}{I}+\kappa)^2}
\end{eqnarray}
and the exponents are $t(\kappa)=1+\kappa$ and $s(\kappa)=1$. In
Fig.~1, we compare for $d=3$ this exact expression for $p_c$ to
numerical results\cite{LeeSiu95,LeeYuen95,Sali97}. It thus seems that
the variational method used in\cite{LeeSiu95,LeeYuen95} does not lead
to the correct values of the electric field around the impurity. The
fact that the percolation threshold depends on $\kappa$ is the bad
feature of this approximation. However, for $\kappa$ not too large, or
for a contrast not too high this approximation works well. When one
adds a {\it linear} background of effective conductivity $\sigma_e$,
an additional dimensionless factor
$\Lambda=\sigma_e/\chi_eE_0^{\kappa}$ is introduced in the
equations. Close to $p_c$, the effective conductivity behaves as
$\sigma_e\sim\Delta p^1$ and the effective nonlinear susceptibility as
$\chi_e\sim\Delta p^{1+\kappa}$. The factor $\Lambda$ is then
diverging and the behavior of $f$ is modified. One then recovers the
Bruggeman value for the percolation threshold
$p_c=1/d$. It thus seems that in the one-impurity
scheme a linear background is necessary to ``regularize'' the wrong
behavior of $p_c$. It might be a way to obtain an EMA satisfying
conditions (i) and (ii), and which gives a correct value for
$p_c$.

Acknowledgements: I would like to thank D.J.~Bergman,
Y.-P.~Pellegrini, P.~Ponte Casta\~neda for many interesting and
stimulating discussions. I also would like to thank J.-M.~Diani,
H. Orland, and G.~Zerah for their interest in this work.




\begin{figure}
\narrowtext
\centerline{
\epsfysize=0.9\columnwidth{\rotate[r]{\epsfbox{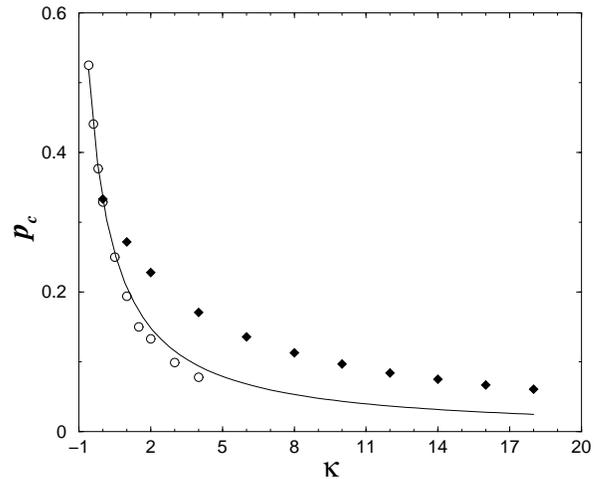}}}
}
\vspace*{1.0cm}
\caption{$p_c$ versus $\kappa$ for $d=3$. The line is the analytical
expression (26), the circles and the diamonds represent the numerical results
from [35] and [25,33] respectively.}
\label{figure2}
\end{figure}


\end{multicols}


\begin{references}


\bibitem{Landauer78} R. Landauer, in {\it Proceedings of the first
Conference on Electrical and Optical Properties of Inhomogeneous
Media} (Ohio State University, 1977), AIP Conf. Proc. No. 40, edited
by J. C. Garland and D. B. Tanner (AIP, New York, 1978).

\bibitem{BergmanStroud92} D.J.~Bergman and D.~Stroud, Solid State Physics
{\bf 46}, 147 (1992).


\bibitem{StroudHui88} D.~Stroud and P.M.~Hui, Phys. Rev. B {\bf 37},
8719 (1988).

\bibitem{Zeng88} X.C.~Zeng, D.J.~Bergman, P.M.~Hui, and D.~Stroud, 
Phys. Rev. B (Rapid Com.) {\bf 38}, 10970 (1988).

\bibitem{Zeng89} X.C.~Zeng, P.M.~Hui, D.J.~Bergman, and D.~Stroud, 
Physica A {\bf 157}, 192 (1989).

\bibitem{Tremblay92} R.R.~Tremblay, G.~Albinet, and A.-M.S.~Tremblay,
Phys. Rev. B {\bf 45}, 755 (1992).

\bibitem{Ponte92} P.~Ponte~Casta\~neda, G.~deBotton, and G.~Li,
Phys. Rev. B {\bf 46}, 4387 (1992).

\bibitem{Zhang94} X. Zhang and D. Stroud, Phys. Rev. B {\bf 49}, 944 (1994).

\bibitem{Levy94} O.~Levy and D.J.~Bergman, Phys. Rev. B {\bf 50}, 3652 (1994).

\bibitem{HuiWan95} P.M.~Hui, W.M.V.~Wan, and K.H.~Chung, Phys. Rev. B
{\bf 52}, 15867 (1995).

\bibitem{Pellegrini00} Y.-P.~Pellegrini, Phys. Rev. B {\bf 61}, 9365 (2000).


\bibitem{Chakrabarti88} B.K.~Chakrabarti, Rev. Solid State Sci.{\bf
2}, 559 (1988).

\bibitem{Roux90} H.~Herrmann and S.~Roux (Eds.) {\it Statistical
Models for the Fracture of Disordered Media}, North-Holland (1990),
and references cited therein.

\bibitem{Einziger87} R.~Einziger, Annu. Rev. Mater. Sci. {\bf 17}, 299 (1987).

\bibitem{Niklasson89} G.A.~Niklasson, Physica A {\bf 157}, 482 (1989).

\bibitem{Osofsky88} M.~Osofsky, M.~LaMadrid, J.-B.~Bieri,
J.~Gavilano, and J.-M.~Mochel, Phys. Rev. B {\bf 38}, 8486 (1988).


\bibitem{Blumenfeld89} R.~Blumenfeld and D.J.~Bergman, Phys. Rev. B
(Rapid Comm.) {\bf 40}, 1987 (1989).

\bibitem{Barthelemy98} M.~Barth\'el\'emy and H. Orland, European
Physical Journal {\bf 6}, 537 (1998).

\bibitem{Levy92} O.~Levy and D.J.~Bergman, Phys. Rev. B {\bf 46}, 7189 (1992).

\bibitem{Hui96} P.M.~Hui and W.M.V.~Wan, Appl. Phys. Lett. {\bf 69},
1810 (1996).

\bibitem{Straley84} J.P.~Straley and S.W.~Kenkel, Phys. Rev. B {\bf
29}, 6299 (1984)

\bibitem{Bergman91} D.J.~Bergman, in {\it Composite Media and
Homogeneization Theory}, edited by G.~dal Maso and G.F.~Dell'Antonio
(Birkhauser, Boston, 1991).

\bibitem{Ponte92b} P.~Ponte~Casta\~neda, Proc. R. Soc. London A {\bf 341}, 
531 (1992).


\bibitem{Wan96} W.M.V.~Wan, H.C.~Lee, P.M.~Hui and K.W.~Yu,
Phys. Rev. B {\bf 54}, 3946 (1996).

\bibitem{LeeSiu95} H.-C.~Lee, W.-H.~Siu, and K.W.~Yu, Phys. Rev. B
{\bf 52}, 4217 (1995).

\bibitem{Meir86} Y.~Meir, R.~Blumenfeld, A.~Aharony, and A.B.~Harris,
Phys. Rev. B {\bf 34}, 3424 (1986).

\bibitem{HuiWoo95} P.M.~Hui, Y.F.~Woo, and W.M.V.~Wan,
J. Phys. Condens. Matter {\bf 7}, L593 (1995).

\bibitem{Gao96} L.~Gao and Z.-Y.~Li, Phys. Letters A {\bf 219}, 324
(1996).

\bibitem{Hui97} P.M.~Hui, P.~Cheung, and Y.R.~Kwong, Physica A {\bf
241}, 301 (1997).

\bibitem{Ponte97} P.~Ponte~Casta\~neda and M.~Kailasam,
Proc. R. Soc. Lond. A {\bf 453}, 793 (1997).

\bibitem{Ponte98} P.~Ponte~Casta\~neda, Phys. Rev. B {\bf 57}, 12077 (1998).

\bibitem{Bruggeman35} D.A.G.~Bruggeman, Ann. Phys. (Leipzig) {\bf
24}, 636 (1935).


\bibitem{LeeYuen95} H.-C.~Lee, K.-P.~Yuen, and K.W.~Yu, Phys. Rev. B
{\bf 51}, 9317 (1995).

\bibitem{LeeYu95} H.-C.~Lee, K.W.~Yu, and G.Q.~Gu, J. Phys.:
Condens. Matter {\bf 7}, 8785 (1995).

\bibitem{Sali97} L.~Sali and D.J.~Bergman, J. Stat. Phys. {\bf 86},
455 (1997). L.~Sali and D.J.~Bergman, J. Stat. Phys. {\bf 89}, 1105
(1997).

\bibitem{PellegriniBarthelemy00} Y.-P.~Pellegrini and
M.~Barth\'el\'emy, Phys. Rev. E {\bf 61}, 3547 (2000).

\end{references}
\end{document}